\documentclass[12pt]{article}

\usepackage{newtxtext,newtxmath}

\usepackage{graphicx}

\usepackage[letterpaper,margin=1in]{geometry}

\renewenvironment{abstract}
	{\quotation}
	{\endquotation}

\date{}

\makeatletter
\renewcommand{\fnum@figure}{\textbf{Figure \thefigure}}
\renewcommand{\fnum@table}{\textbf{Table \thetable}}
\makeatother

\usepackage{scicite}

\usepackage{url}

\def\scititle{
	Measurement of the $\Lambda$ electric dipole moment with an entangled baryon-antibaryon system
}
\title{\bfseries \boldmath \scititle}

\author{
	BESIII Collaboration
    \thanks{BESIII Collaboration collaborators and affiliations are listed in the supplementary materials.}
}

\begin{document} 

\maketitle

\begin{abstract} \bfseries \boldmath
The dominance of matter over antimatter in the universe has consistently driven the pursuit of new physics beyond the Standard Model that violates charge-parity symmetry.
Strange baryons (hyperons) remain a largely unexplored territory in which interactions between hyperons and particles from new physics could induce a nontrivial electric dipole moment (EDM).
However, direct measurements of hyperon EDMs through spin precession are highly challenging owing to their short lifetimes. In this work, we introduce a method to extract the EDM of the lightest hyperon, $\Lambda$, using the entangled $\Lambda$$\overline{\Lambda}$ system. Our result is consistent with zero, achieving a three-orders-of-magnitude improvement over the previous upper limit established in the 1980s with comparable statistics, providing stringent constraints on potential new physics.

\end{abstract}

\noindent
The existence of a nonzero electric dipole moment (EDM) in a particle signals the violation of time-reversal (T) symmetry and, consequently, charge-parity (CP) symmetry, assuming the combined CPT symmetry holds. 
The amount of CP violation (CPV) in the Standard Model (SM), arising from a complex phase in the quark mixing matrix and the quantum chromodynamics (QCD) vacuum angle $\overline{\theta}$, is insufficient to explain the domination of matter over antimatter in the universe~\cite{Sakharov:1967dj, Riotto:1998bt}.
The measurement of EDMs, which are highly sensitive probes for physics beyond the SM (BSM) in the multi-\(10^{14}\) electron volt mass range, has been conducted across a wide variety of particles, atoms, and molecules~\cite{ACME:2013pal, ACME:2018yjb, Roussy:2022cmp, Abel:2020pzs, Muong-2:2008ebm, Belle:2021ybo, Pondrom:1981gu, Murthy:1989zz, Graner:2016ses, Graner:2017erratum}. 
However, no EDM signal has been observed experimentally to date.
The underlying interactions can be expressed by the effective Lagrangian~\cite{Chupp:2017rkp} 
\begin{align}
L=-\frac{\alpha_s}{8\pi} \bar{\theta} \mathrm{Tr}[G^{\mu\nu}\tilde{G}_{\mu\nu}]-\frac{i}{2}\sum_q d_q \bar{q} F_{\mu\nu}\sigma^{\mu\nu}\gamma^5 q 
-\frac{i}{2}\sum_q \tilde{d}_q \bar{q} G^a_{\mu\nu} T^a\sigma^{\mu\nu}\gamma^5 q,
\label{eq:L}
\end{align}
which encodes the contributions to CP violation and EDMs in the SM and beyond. 
Here, $\alpha_s$ is the strong coupling constant, $F_{\mu\nu}$ and $G_{\mu\nu}=G_{\mu\nu}^{a}T^{a}$ represent the strength of electromagnetic and gluonic fields, respectively. 
The parameters $d_q$ and $\tilde{d}_q$ correspond to the EDM and chromo-EDM (cEDM) of the quarks, respectively, with the latter arising from CP-violating interactions between the quarks and gluons.
The parameters in Eq.~\ref{eq:L} can be constrained from the measurements of EDM in different systems. 
Systematic searches in baryon systems could provide insights into the QCD vacuum and, in conjunction with leptons~\cite{ACME:2013pal, ACME:2018yjb, Roussy:2022cmp}, are essential for uncovering BSM physics~\cite{Beacham:2019nyx}.
An unexpectedly large EDM resulting from interactions between strange quarks in hyperons and new particles could indicate specific types of BSM physics. To achieve a comprehensive understanding of the fundamental laws of physics, a global analysis of EDM searches across various systems, simultaneously constraining the parameters of a general theory integrating SM and BSM physics, requires experimental input from hyperons, baryons containing at least one strange quark~\cite{Pospelov:2005pr, Chupp:2017rkp}.

\begin{figure}[!htbp]
        \centering
        \includegraphics[width=0.95\textwidth]{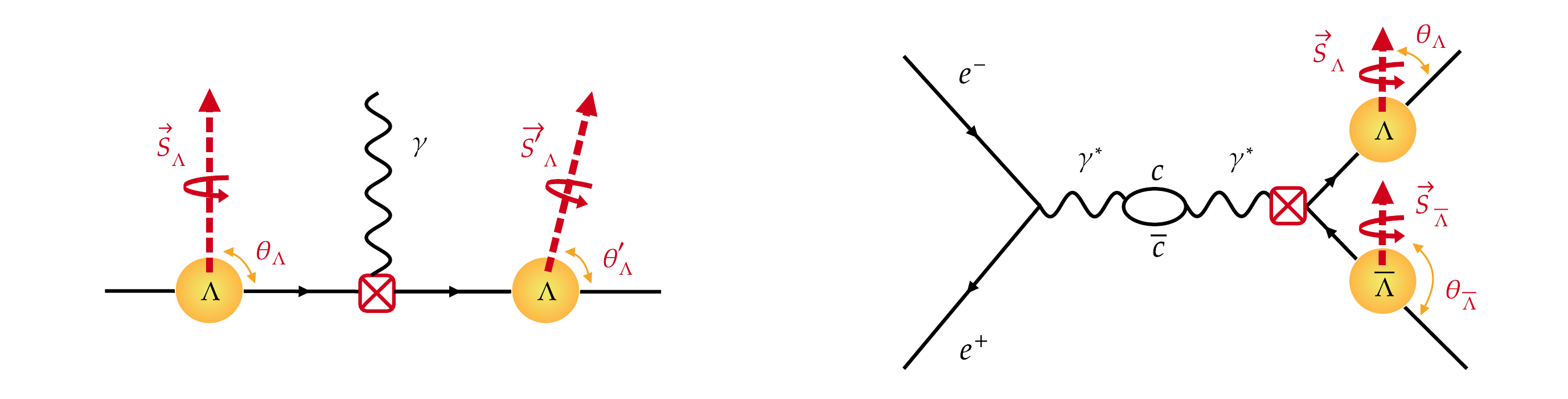} 
        \makebox[0pt][l]{\hspace*{-36em}\raisebox{9em}[0pt][0pt]{\fontfamily{phv}\selectfont\bfseries\fontsize{10}{12}\selectfont A}}%
        \makebox[0pt][l]{\hspace*{-18em}\raisebox{9em}[0pt][0pt]{\fontfamily{phv}\selectfont\bfseries\fontsize{10}{12}\selectfont B}}%
    \caption{\textbf{Schematic view of EDM measurements.} (\textbf{A}) Determination through spin precession induced by an external electromagnetic field. Here, $\vec{S}_{\Lambda}$ and $\vec{S}_{\Lambda}^{\prime}$ denote the spin directions of the $\Lambda$ particle before and after precession, whereas $\theta_{\Lambda}$ and $\theta_{\Lambda}^{\prime}$ represent the angles between the spin and momentum directions before and after precession.  
    (\textbf{B}) Extraction through a CP-violating form factor induced by virtual-photon exchange between the $J/\psi$ and the polarized, spin-entangled $\Lambda\overline{\Lambda}$ system. Here, $\vec{S}_{\Lambda}$ and $\vec{S}_{\overline{\Lambda}}$ are the spin directions of the $\Lambda$ and $\overline{\Lambda}$, respectively, whereas $\theta_{\Lambda}$ and $\theta_{\overline{\Lambda}}$ represent the angles between their spin and momentum directions. The red, \scalebox{0.6}{$\boxed{\times}$} symbol indicates a CP-violating interaction vertex. $\gamma$ and $\gamma^{*}$ denote a photon and a virtual photon, respectively.}
\label{fig:edm_diagram}
\end{figure}

The interaction between the spin of a hyperon and an external electromagnetic field induces a precession effect, which can be used to directly measure the hyperon’s EDM (Fig.~\ref{fig:edm_diagram}A). This approach has been widely used for EDM measurements of electrons, neutrons, and protons but remains challenging for short-lived particles, such as hyperons. In addition to their short lifetime, producing a well-polarized particle source is difficult.
To date, the \(\Lambda\), the lightest hyperon, is the only hyperon whose EDM has been measured; no other EDM measurements have been conducted in the hyperon sector, although some have been proposed~\cite{Botella:2016ksl, Bagli:2017foe, Bagli:2020erratum}.
The most stringent upper limit on the \(\Lambda\) hyperon EDM, \(1.5 \times 10^{-16}\) \(e\) cm, was established in 1981 at Fermilab~\cite{Pondrom:1981gu}. This study also highlighted several challenges that limited further improvements in sensitivity. Although increasing the signal yield, optimizing the \(\Lambda\) polarization, and extending the effective magnetic field coverage could enhance precision, reference~\cite{Pondrom:1981gu} suggested that the spin-precession technique would be difficult to push beyond the \(10^{-18}\) \(e\) cm level.

\clearpage

In this study, we present an alternative approach in which the spin-entangled \(\Lambda\)\(\overline{\Lambda}\) pair is produced through the process \(e^+e^- \rightarrow J/\psi \rightarrow \Lambda \overline{\Lambda}\), where $J/\psi$ is a charmonium state composed of a charm quark and an anticharm quark. 
The subtle difference between the angular distributions of the $\Lambda$ and $\overline{\Lambda}$ can be used to access the EDM. 
Within the SM, the interaction between the  $J/\psi$ and the $\Lambda\overline{\Lambda}$ pair involves the strong force, quantum electrodynamics, and the weak force. The dominant contribution comes from nonperturbative QCD.
Because the \(\Lambda\) is not a point-like particle, its internal structure and the aforementioned interactions are encapsulated in form factors, which depend on the production energy of the \(\Lambda\overline{\Lambda}\). The violation of parity and CP symmetry in \(\Lambda\overline{\Lambda}\) pair production is characterized by the form factors \( F_A \) and \( H_T \), respectively, and can originate from various sources. Here, the subscripts $A$ and $T$ label the axial and tensor structures. 
As discussed in~\cite{He:1992ng, He:1993ar, He:2022jjc, Du:2024jfc}, the form factor $F_A$ arises from the parity-violating decay of the $J/\psi$ through intermediate $Z$- and $W$-boson exchanges, whereas the CP-violating form factor $H_T$ is expected to be dominated by contributions from the $\Lambda$ EDM through virtual-photon exchange between the $J/\psi$ and the $\Lambda\overline{\Lambda}$ pair (Fig.~\ref{fig:edm_diagram}B).
The form factor \( H_T \) can be related to the \(\Lambda\) EDM (\( d_{\Lambda} \)) while neglecting its dependence on the production energy~\cite{Du:2024jfc}, 
\begin{equation}
	\begin{aligned}
        H_{T}=\frac{2e}{3m^{2}_{J/\psi}}g_{V}d_{\Lambda}.
	\end{aligned}
\end{equation}
Here, $e$ denotes the magnitude of the electron charge, $g_V$ is determined from the measured branching fraction of $J/\psi\to e^+e^-$~\cite{He:1992ng}, and $m_{J/\psi}$ 
represents the mass of $J/\psi$.
Parity symmetry is violated in both the production and decay of the \( J/\psi \) meson owing to weak force contributions. In the production process, the weak interaction in electron-positron annihilation induces a small longitudinal polarization, \( P_L \). The polarization \( P_L \) and the parity-violating form factor \( F_A \) are predicted to be of the order of \( \mathcal{O}(10^{-4}) \) and \( \mathcal{O}(10^{-6}) \), respectively~\cite{He:2022jjc,Du:2024jfc}.

The extraction of $d_{\Lambda}$ through the CP-violating form factor \( H_T \) in \( J/\psi \to \Lambda\overline{\Lambda} \) decays was first proposed in~\cite{He:1992ng, He:1993ar}. 
We implemented this idea in our experiment by using the process $e^+e^- \rightarrow J/\psi \rightarrow \Lambda \overline{\Lambda}$, with $\Lambda(\overline{\Lambda})$ decaying to proton $p$ (antiproton $\bar{p}$) and charged pion $\pi^-$ ($\pi^+$).
This approach~\cite{Fu:2023ose} utilizes the differential decay probability, derived from the full angular distributions of the production and decay of the spin-entangled $\Lambda \overline{\Lambda}$ system. 
The entanglement, arising from angular momentum conservation, enables a highly sensitive probe of the EDM. 
The EDM effect, which manifests as a difference in the angular distributions of entangled \(\Lambda\) and \(\overline{\Lambda}\) baryons, can be visualized through the projection of the triple-product asymmetry (TPA), $O \equiv (\hat{l}_{p} \times \hat{l}_{\overline{p}}) \cdot \hat{k}$, which is proportional to the real part of the form factor $H_T$~\cite{He:2022jjc,Du:2024jfc}. In this case, $\hat{l}_{p}(\hat{l}_{\overline{p}})$ and $\hat{k}$ represent the unit momentum of the proton (antiproton) in the rest frame of the $\Lambda(\overline{\Lambda})$ and the unit momentum of $\Lambda$ in the center-of-mass frame of the $e^{+}e^{-}$ system, respectively.

\subsection*{Experimental overview}

\begin{figure}[!htbp]
    \centering
    \includegraphics[width=0.9\textwidth]{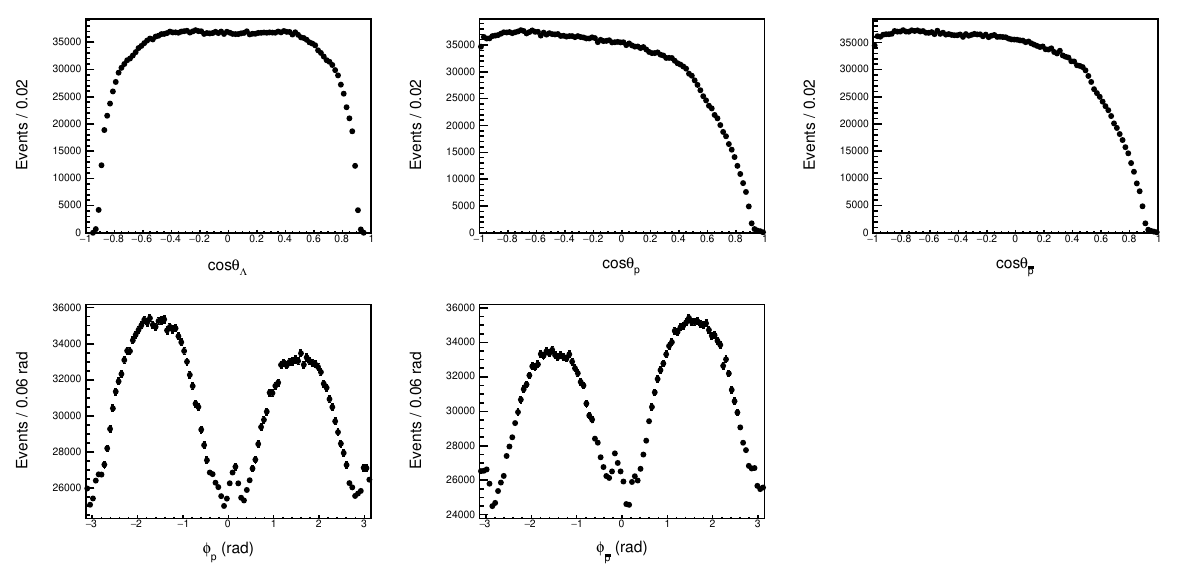}
    \makebox[0pt][l]{\hspace*{-34em}\raisebox{17em}[0pt][0pt]{\fontfamily{phv}\selectfont\bfseries\fontsize{10}{12}\selectfont A}}%
    \makebox[0pt][l]{\hspace*{-22.3em}\raisebox{17em}[0pt][0pt]{\fontfamily{phv}\selectfont\bfseries\fontsize{10}{12}\selectfont B}}%
    \makebox[0pt][l]{\hspace*{-10.5em}\raisebox{17em}[0pt][0pt]{\fontfamily{phv}\selectfont\bfseries\fontsize{10}{12}\selectfont C}}%
    \makebox[0pt][l]{\hspace*{-34em}\raisebox{8.5em}[0pt][0pt]{\fontfamily{phv}\selectfont\bfseries\fontsize{10}{12}\selectfont D}}%
    \makebox[0pt][l]{\hspace*{-22.3em}\raisebox{8.5em}[0pt][0pt]{\fontfamily{phv}\selectfont\bfseries\fontsize{10}{12}\selectfont E}}%
    \caption{\textbf{Distributions of the five kinematic variables used in the analysis.}
  \textbf{(A)} \(\cos\theta_{\Lambda}\),
  \textbf{(B)} \(\cos\theta_{p}\),
  \textbf{(C)} \(\cos\theta_{\bar{p}}\),
  \textbf{(D)} \(\phi_{p}\),
  and \textbf{(E)} \(\phi_{\bar{p}}\).
  Points with error bars represent the experimental data.
  \(\theta_{\Lambda}\) is the polar angle of the \(\Lambda\) hyperon with respect to the \(e^{-}\) beam direction in the \(e^{+}e^{-}\) center-of-mass frame. \(\theta_{p}\) (\(\theta_{\bar{p}}\)) and \(\phi_{p}\) (\(\phi_{\bar{p}}\)) are the polar and azimuthal angles of the proton (antiproton) in the rest frame of the \(\Lambda\) (\(\overline{\Lambda}\)).}
    \label{fig:angles}
\end{figure}

We selected 3 million high-purity \(\Lambda\overline{\Lambda}\) pairs from a dataset of 10 billion \( J/\psi \) events collected at BESIII in 2009, 2012, and 2017 to 2019~\cite{BESIII:2021cxx}.
The signal events were selected through a stringent procedure: Charged particle tracks were reconstructed in the main drift chamber, where a superconducting solenoid generated a magnetic field that enabled precise momentum measurement, achieving an accuracy of $0.5\%$ at 1.0 GeV/$c$. 
Protons and charged pions were identified based on their momentum information, with high-momentum tracks classified as proton candidates, and low-momentum tracks classified as pion candidates. 
The \(\Lambda(\overline{\Lambda})\) candidates were formed using $p\pi^-(\bar{p}\pi^+)$ pairs, with a vertex fit ensuring a common decay position. Their reconstructed masses are required to lie within the \(\Lambda\) mass window, specifically between 1.111 and \(1.121~\text{GeV}/c^{2}\).
A kinematic fit imposing energy-momentum conservation constraint improved the momentum resolution of the final-state particles.
 After applying selection criteria, the final dataset consisted of 3,049,187 signal candidates with a purity of \( 99.9\% \). 
For each event, we computed a complete set of kinematic variables, denoted as \( \boldsymbol{\xi} = (\theta_{\Lambda}, \theta_{p}, \theta_{\bar{p}}, \phi_{p}, \phi_{\bar{p}}) \), using the momenta of both intermediate and final-state particles. In this case, \( \theta_{\Lambda} \) represents the angle between the \( \Lambda \) hyperon and the electron beam in the center-of-mass system of \( e^+e^- \), whereas \( \theta_{p} \) and \( \phi_{p} \) (\( \theta_{\bar{p}} \) and \( \phi_{\bar{p}} \)) correspond to the polar and azimuthal angles of the proton (antiproton) in the rest frame of the \( \Lambda(\overline{\Lambda}) \). 
The distributions of these angular variables in the experimental data are shown in Fig.~\ref{fig:angles}.

For the two-body decay \(\Lambda \to p\pi\), these proton (antiproton) angular variables provide a complete description of the decay kinematics, so using the proton information alone fully captures the relevant angular distribution.
We then constructed the likelihood function using the differential decay probability of $e^+e^- \rightarrow J/\psi \rightarrow \Lambda (\rightarrow p\pi^-)\overline{\Lambda}(\rightarrow \bar{p}\pi^+)$ as a function of \( \xi \) given by eq.~\ref{angdis} in \cite{methods}. The longitudinal polarization $P_L$, form factors \( H_T \) and \( F_A \), and other parameters that are defined in \cite{methods} were extracted by minimizing the negative log-likelihood function, accounting for both data and background contribution. 
During data analysis, the values of parameters $P_{L}$, $H_{T}$, and $F_{A}$ were blinded by introducing unknown offsets, which were not removed until the event selection, fitting procedure, and systematic uncertainty estimation were fully completed. Prior to unblinding, the full analysis chain was validated using simulated samples generated under the null EDM hypothesis for which no shift in the EDM was observed. In addition, the experimental data were divided into subsamples according to data-taking period, $\Lambda$ polar angle in the center-of-mass frame, and $\Lambda/\bar{\Lambda}$ decay length. 
For each subsample, the same unknown offset as in the nominal fit was applied, and only the deviations from the nominal result were examined. In all cases, these deviations were consistent with zero within statistical uncertainties, showing no evidence of systematic shifts in the EDM measurement. A comprehensive description of the above procedure has been provided in \cite{methods}.

\begin{table}[!htbp]
    \centering
    \caption{
    \textbf{Summary of fitted parameters.} 
   The extracted parameters include the \( J/\psi \) polarization \( P_L \), which characterizes parity violation in \( J/\psi \) production, as well as the form factors \( F_A \) and \( H_T \), which represent parity and CP violation in \( J/\psi \) decay, respectively. The \(\Lambda\) electric dipole moment (\( d_{\Lambda} \)) was derived from the fitted value of $H_{T}$. The first uncertainty represents the statistical uncertainty, whereas the second corresponds to the systematic uncertainty.}
    \scalebox{1.0}{
        \begin{tabular}{ll}
            \\
            \hline
            \hline
            \textbf{\textrm{Parameter}} & \textbf{\textrm{Fitting results}}  \\ 
            \hline 
            $P_{L}$      & $(-1.8  \pm 1.2 \pm 0.8)   \times 10^{-3}$    \\ \hline 
            $Re(F_{A})$  & $(-2.4  \pm 1.6 \pm 3.1)   \times 10^{-6}$    \\ \hline 
            $Im(F_{A})$  & $(-7.9  \pm 3.7 \pm 2.5)   \times 10^{-6}$  \\ \hline 
            $Re(H_{T})$  & $(-1.4  \pm 1.4 \pm 0.2)   \times 10^{-6}$ $\textrm{GeV}^{-1}$    \\ \hline 
            $Im(H_{T})$  & $(1.3   \pm 1.2 \pm 0.4)   \times 10^{-6}$ $\textrm{GeV}^{-1}$    \\  
            \hline 
            \hline 
            $Re(d_{\Lambda})$  & $(-3.1 \pm 3.2 \pm 0.5) \times 10^{-19}$ $e\ \textrm{cm}$ \\ \hline
            $Im(d_{\Lambda})$  & $(2.9 \pm 2.6 \pm 0.6) \times 10^{-19}$ $e\ \textrm{cm}$  \\ \hline
            \hline
        \end{tabular}
    }
    \label{tab:fit_results}
\end{table}
\clearpage

\subsection*{Systematic uncertainty evaluation}
The potential sources of systematic uncertainty in this measurement include background estimation, charged particle reconstruction, vertex reconstruction for $\Lambda(\overline{\Lambda})$, energy–momentum conservation constraints, and detector resolution. A comprehensive summary of these uncertainties is provided in table~\ref{tab:sys_summary}, and further details of the systematic studies can be found in \cite{methods}. Among these, the contributions from background estimation and charged particle reconstruction were evaluated and found to be negligible.
The systematic uncertainties related to the $\Lambda$ and $\overline{\Lambda}$ vertex reconstruction and the energy–momentum conservation constraints were evaluated using a data-driven study of the efficiency differences between data and simulation. The nominal fit was performed without applying efficiency corrections; the differences observed when applying the efficiency corrections were taken as the associated systematic uncertainty.
The systematic uncertainty arising from detector resolution was evaluated using simulated data resampling according to the differential angular probability shown in eq.~\ref{angdis}. The discrepancy between the fitted and the generated values was taken as the systematic uncertainty.

To further evaluate the accuracy of our measurement, we systematically investigated potential sources of shifts in the EDM, including imperfections in the simulated data and maximum likelihood estimator, as well as \(\Lambda\)($\overline{\Lambda}$) decay vertex reconstruction and corrections for final-state radiation (FSR) from charged final-state particles.  
The channel of $J/\psi \rightarrow p\pi^- \bar{p}\pi^+$ provides reliable efficiency for charged particle reconstruction and allows us to apply corrections to the simulated data. 
A set of pseudoexperiments was generated under the null EDM hypothesis to evaluate the behavior of the maximum likelihood estimator. The pull distribution, defined as the difference between the fitted results and the null EDM hypothesis divided by the corresponding measurement uncertainty, is consistent with a normal distribution, and no shift in the EDM was observed. 
Additionally, we examined whether the reconstruction of the \(\Lambda\)($\overline{\Lambda}$) decay vertex could introduce a potential shift. The data sample was split into several groups according to the \(\Lambda\)($\overline{\Lambda}$) decay length to evaluate the consistency of the measurements, and no shift was observed. 
The effect of FSR was evaluated with simulated data generated using the PHOTOS Monte Carlo program~\cite{Richter-Was:1992hxq}. The nominal fit used the simulation data, including FSR, in $\Lambda$ decays, and the difference with the fit using a sample without FSR was taken as the systematic uncertainty, which was found to be negligible.

\subsection*{EDM results}
The results of the fitted parameters are summarized in Table~\ref{tab:fit_results}. The measurements of the longitudinal polarization of the \( J/\psi \) resonance $P_L$ and parity-violating form factor \( F_A \) approached the theoretical prediction limits~\cite{Du:2024jfc}. The measurement of the \(\Lambda\) EDM was consistent with zero, preserving CP symmetry with a precision of \( 10^{-19} \) \( e \) cm,
\begin{equation}
    \begin{aligned}
        Re(d_{\Lambda}) &= (-3.1 \pm 3.2 \pm 0.5) \times 10^{-19}\ e\ \rm{cm}, \\
        Im(d_{\Lambda}) &= (2.9 \pm 2.6 \pm 0.6) \times 10^{-19}\ e\ \rm{cm}.
    \end{aligned}
\end{equation}

The upper limits on the \(\Lambda\) EDM were obtained from a profile likelihood scan of the real and imaginary components of \(H_T\) using the RooFit framework~\cite{Verkerke:2003ir}. The resulting 95\% confidence level (CL) intervals for \(\mathrm{Re}(H_T)\) and \(\mathrm{Im}(H_T)\) were then translated into the corresponding limits on \(\mathrm{Re}(d_{\Lambda})\) and \(\mathrm{Im}(d_{\Lambda})\). The extracted 95\% CL intervals are  
\begin{equation}
    \begin{aligned}
       -8.6 \times 10^{-19} < \mathrm{Re}(d_{\Lambda}) < 3.3 \times 10^{-19}\ e\,\text{cm}, \\
       -2.5 \times 10^{-19} < \mathrm{Im}(d_{\Lambda}) < 7.2 \times 10^{-19}\ e\,\text{cm}.
    \end{aligned}
\end{equation}
We used the CL\(_{\mathrm{s}}\) method~\cite{Read:2002hq} to derive the upper limit on \(|H_T|\), which was then translated into an upper bound on the magnitude of

\begin{equation}
	\begin{aligned}
        |d_{\Lambda}| < 7.0 \times 10^{-19}~e~\text{cm} \quad \text{(95\% CL)} .
	\end{aligned}
\end{equation}

The visualization of the effect of EDM was performed by projecting the difference between the positive and negative values of TPA, 
\( A(O) = \frac{N_{\text{event}}(O>0)-N_{\text{event}}(O<0)}{N_{\text{event}}(O>0)+N_{\text{event}}(O<0)} \). 
The variation of the CP asymmetry \(A(O)\) with \(|O|\) was demonstrated using simulated events resampled from the nominal model corresponding to the fit results of the data.
To demonstrate the impact of a nonzero EDM, which may be allowed in certain BSM models, we also considered an alternative scenario in which \( H_T \) deviates from zero by 10 standard deviations, corresponding to $|d_{\Lambda}| = 4.2\times10^{-18}\ e$ cm. The EDM visualization for both scenarios is shown in Fig.~\ref{fig:combined_edm}A, highlighting the differences between the measurement in this analysis and the nonzero EDM scenario.

\subsection*{Discussion and outlook}

\begin{figure*}[!htbp]
    \centering

    \begin{minipage}{0.52\textwidth}
        \centering
        \includegraphics[width=\textwidth]{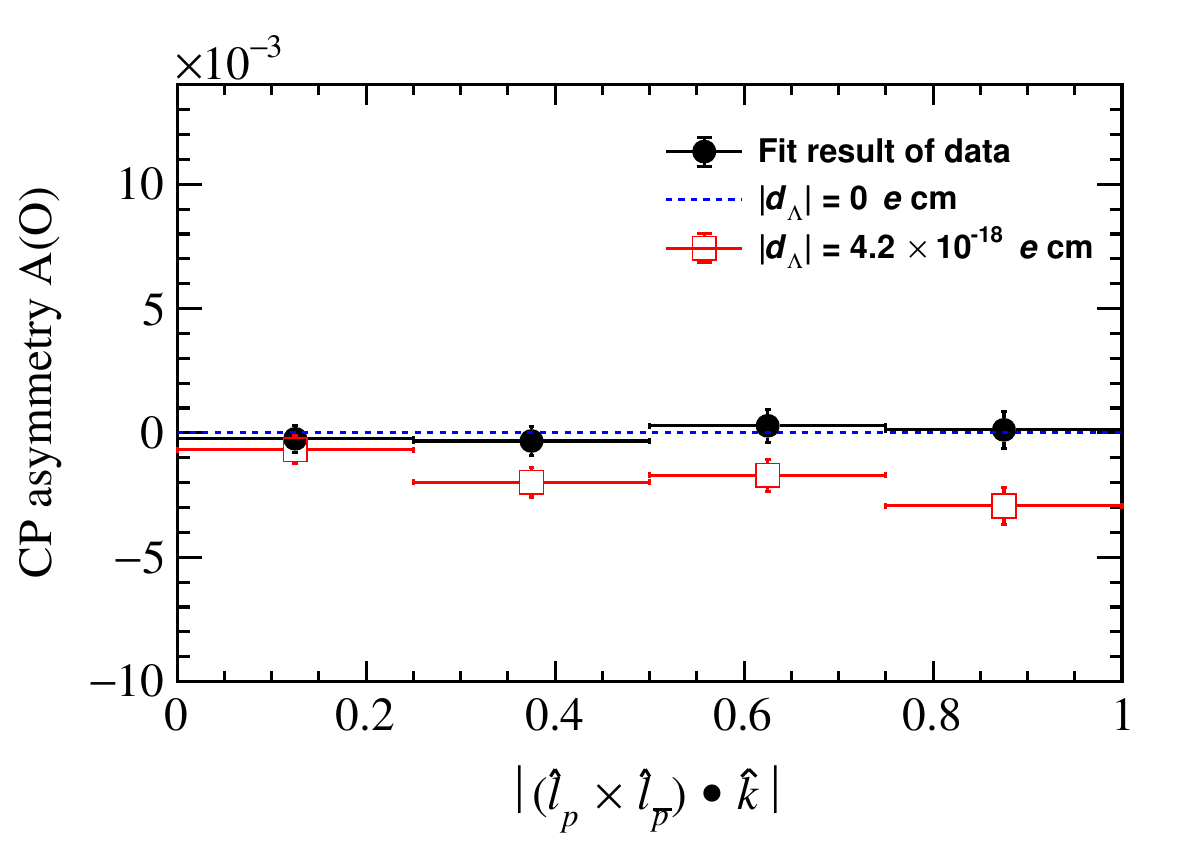}
        \makebox[0pt][l]{\hspace*{-6em}\raisebox{13em}[0pt][0pt]{\fontfamily{phv}\selectfont\bfseries\fontsize{10}{12}\selectfont A}}%
    \end{minipage}

    \vspace{0.8em}

    \begin{minipage}{0.4\textwidth}
        \centering
        \includegraphics[width=\textwidth]{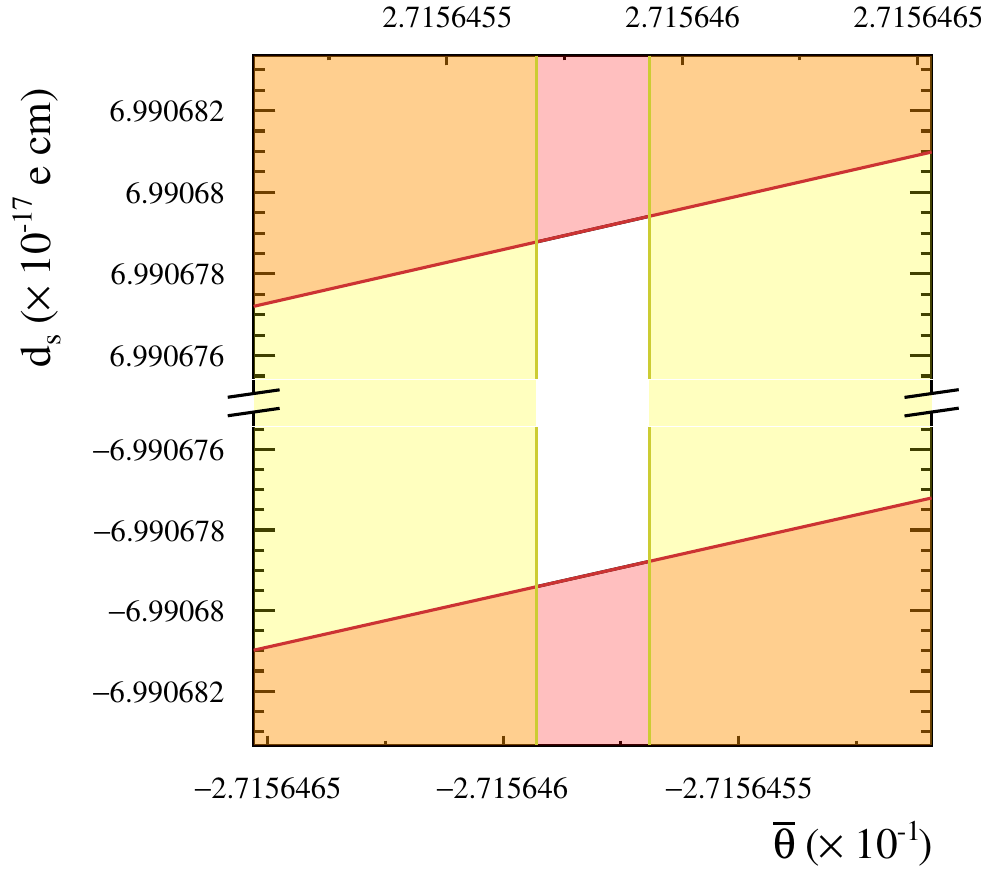}
        \makebox[0pt][l]{\hspace*{-2.5em}\raisebox{13em}[0pt][0pt]{\fontfamily{phv}\selectfont\bfseries\fontsize{10}{12}\selectfont B}}%
    \end{minipage}
    \hspace{0.03\textwidth}
    \begin{minipage}{0.54\textwidth}
        \centering
        \includegraphics[width=\textwidth]{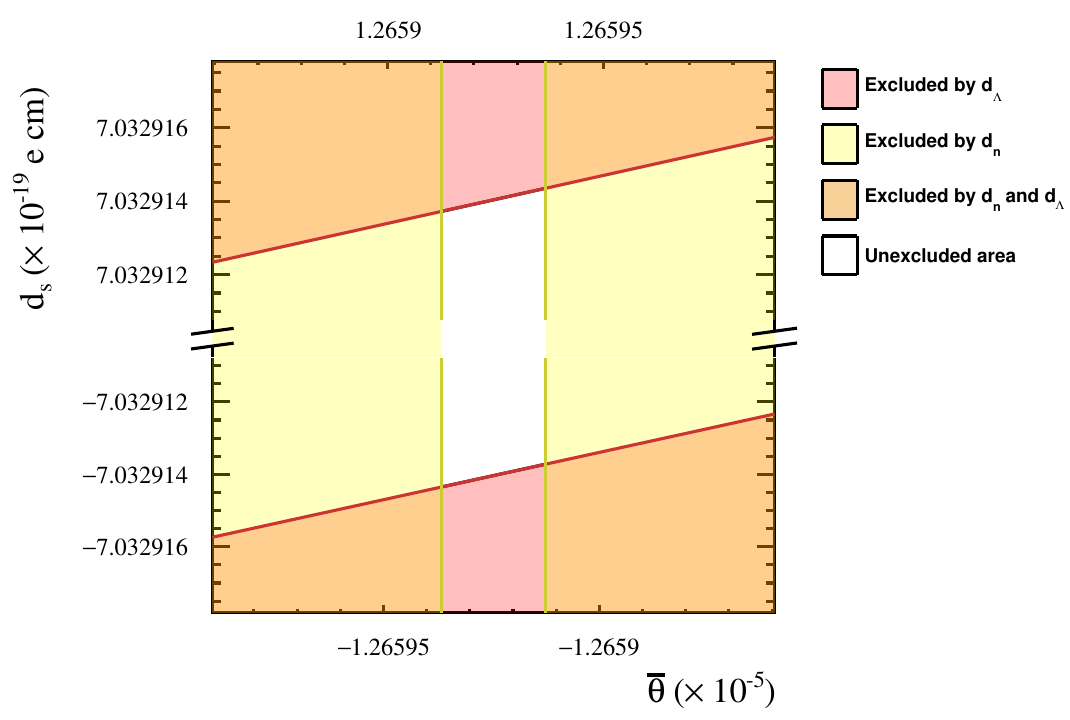}
        \makebox[0pt][l]{\hspace*{-4.7em}\raisebox{13.3em}[0pt][0pt]{\fontfamily{phv}\selectfont\bfseries\fontsize{10}{12}\selectfont C}}%
    \end{minipage}

    \caption{\textbf{Visualization of the EDM effect and constraints on the parameter space of $\overline{\theta}$ and $d_s$.} 
    \textbf{(A)} Projection of the CP asymmetry used to visualize the EDM effect. The black data points represent the projection based on the nominal fit to the data, where \(H_T\) is consistent with zero within uncertainties. The red data points correspond to the projection under a scenario with a significant nonzero EDM, where \(H_T\) deviates from zero by 10 standard deviations, corresponding to $|d_{\Lambda}| = 4.2\times10^{-18}\ e$ cm. The blue dashed line shows the projection assuming a strict zero EDM.
    \textbf{(B and C)} Constraints on the parameter space of $\overline{\theta}$ and $d_s$ for scenarios with \textbf{(B)} and without \textbf{(C)} $SU(3)$ flavor symmetry, respectively. Note that the units of the x and y axes differ between \textbf{(B)} and \textbf{(C)}. Breaks are introduced in the y axes due to their large disparity. The pink region denotes the parameter space excluded by $\Lambda$ EDM measurements, the yellow denotes that excluded by the current upper limit on the neutron EDM, $|d_n| < 1.8\times10^{-26}\ e\ \mathrm{cm}$~\cite{Abel:2020pzs}, and the orange denotes that excluded by their combined constraints. The unshaded region denotes the parameter space that remains unconstrained.}
    \label{fig:combined_edm}
\end{figure*}

In the valence quark model, neglecting contributions from sea quarks, the $\Lambda$ EDM is related to strange quark EDM as $d_\Lambda=d_s$~\cite{Ellis:1989af}. When incorporating the contribution from the QCD vacuum angle $\overline{\theta}$~\cite{Guo:2012vf}, this can be expressed as
\begin{equation}
    \begin{aligned}
        d_{\Lambda}&= (-2.6\pm0.4)\times10^{-16} \ \overline{\theta} ~e~\text{cm} + d_s. 
    \end{aligned}
    \label{eq:d_Lambda}
\end{equation}
On the other hand, the neutron EDM, $d_n$, is given by lattice QCD (LQCD), as in references~\cite{Dragos:2019oxn,Gupta:2018lvp}, where $d_u$ and $d_d$ are the up- and down-quark EDMs, respectively: 
\begin{equation}
    \begin{aligned}
        d_n&=-(1.5\pm0.7)\times 10^{-16}\ \overline{\theta} ~e~\text{cm} \\
        &-(0.20\pm0.01)d_u + (0.78\pm0.03)d_d + (0.0027\pm0.0016)d_s.
    \end{aligned}
    \label{eq:d_n}
\end{equation}
Because the contribution of quark cEDMs remains under investigation by several LQCD groups~\cite{Kim:2021qae}, we did not include these contributions in the present analysis.
As seen from Eqs.~\ref{eq:d_Lambda} and \ref{eq:d_n}, the measurement of the EDM for a single hadron is insufficient to disentangle the various potential sources of CP violation in the quark sector.
Aside from the neutron EDM, the hyperon EDM provides valuable insight into the study of the strange quark EDM.
However, it is still insufficient to independently extract constraints on new physics models. In this work, the measurement further constrains the fundamental parameter space. Alongside the upper limit on the neutron EDM, $|d_n| < 1.8 \times 10^{-26}~e~\text{cm}$~\cite{Abel:2020pzs}, two scenarios were explored.
First, a physics model incorporating \( SU(3) \) flavor symmetry, which indicates that the $u$, $d$, and $s$ quarks behave similarly under strong interactions, is considered, implying \( d_u = d_d = d_s \).
By contrast, given that $m_s>m_u\sim m_d$, a second scenario explores a model without $SU(3)$ flavor symmetry, where $d_s\gg d_u,d_d$. Thus, the contributions of the $u$ and $d$ quarks can be neglected.
The exclusion for the parameters $\overline\theta$ and $d_s$ in these two scenarios are illustrated in Fig.~\ref{fig:combined_edm}, B and C, where the second scenario without $SU(3)$ flavor symmetry provides a more stringent constraint.

A full angular analysis of the $J/\psi\to\Lambda\overline{\Lambda}$ decay was performed, and the EDM of the $\Lambda$ was determined with a precision of $10^{-19}$ $e$ cm, representing a three-orders-of-magnitude improvement over the previous result obtained 40 years ago with comparable statistics~\cite{Pondrom:1981gu}.
Our result, together with the upper limit on the neutron EDM, provides more stringent constraints on the new physics parameter space. 
The method used for the $\Lambda$ EDM measurement can be extended to systematic studies of the hyperon family at BESIII. Simulated data indicate that a comparable precision can be achieved for the $\Xi$ and $\Sigma$ hyperons~\cite{Fu:2023ose}. The present measurement is primarily limited by statistical uncertainties. At future high-luminosity facilities, such as the proposed Super Tau-Charm Factory~\cite{Achasov:2023gey}, substantially larger data samples could improve the sensitivity to the $\Lambda$ EDM by one to two orders of magnitude~\cite{Fu:2023ose}.

\clearpage 

%
\bibliography{LambdaEDM_science} 
\bibliographystyle{sciencemag}

%
%
%
%
%
%


\section*{Acknowledgments}
The author gratefully acknowledges the valuable discussions and theoretical support provided by X.-G. He, J.-P. Ma, and J.-P. Wang. 
The BESIII Collaboration thanks the staff of BEPCII (https://cstr.cn/31109.02.BEPC) and the IHEP computing center for their strong support. \textbf{Funding:} This work is supported in part by the National Key R$\&$D Program of China under contracts nos. 2023YFA1606000, 2020YFA0406300, 2020YFA0406400, and 2023YFA1606704; the National Natural Science Foundation of China (NSFC) under contracts nos. 11635010, 11935015, 11935016, 11935018, 12025502, 12035009, 12035013, 12061131003, 12192260, 12192261, 12192262, 12192263, 12192264, 12192265, 12221005, 12225509, 12235017, 12275058, 12361141819, and 12475082; the Chinese Academy of Sciences (CAS) Large-Scale Scientific Facility Program under contract no. YSBR-101; 100 Talents Program of CAS; The Institute of Nuclear and Particle Physics (INPAC) and Shanghai Key Laboratory for Particle Physics and Cosmology; the Natural Science Foundation of Shan-dong Province under contract no. ZR2023MA004; Agencia Nacional de Investigación y Desarrollo de Chile (ANID), Chile, under contract no. ANID PIA/APOYO AFB230003; the German Research Foundation DFG under contract no. FOR5327; Istituto Nazionale di Fisica Nucleare, Italy; the Knut and Alice Wallenberg Foundation under contract nos. 2021.0174 and 2021.0299; the Ministry of Development of Turkey under contract no. DPT2006K-120470; the National Research Foundation of Korea under contract no. NRF-2022R1A2C1092335; the National Science and Technology fund of Mongolia; the National Science Research and Innovation Fund (NSRF) through the Program Management Unit for Human Resources $\&$ Institutional Development, Research, and Innovation of Thailand under contract no. B50G670107; the Polish National Science Centre under contract no. 2024/53/B/ST2/00975; the Swedish Research Council under contract no. 2019.04595; and the US Department of Energy under contract no. DE-FG02-05ER41374.
\paragraph*{Author contributions:}
All authors have contributed to the publication, being variously involved in the design and the construction of the detectors, writing software, calibrating subsystems, operating the detectors, acquiring data, and analyzing the processed data.
\paragraph*{Competing interests:}
The authors declare no competing interests.
\paragraph*{Data, code, and materials availability:} The data underlying the plots in this paper are available on Zenodo~\cite{zenodo:2025}. The algorithms used for data analysis are available on Zenodo~\cite{zenodo_code:2026}.


\subsection*{Supplementary materials}
BESIII Collaboration Author List\\
Materials and Methods\\
Fig.~S1 \\
Tables~S1 to S3\\
References~\cite{BESIII:2009fln,BESIII:2020nme,GEANT4:2002zbu,Deng:2006,Jadach:2000ir,Ping:2008zz,Chen:2000tv,Yang:2014vra,Liao:2009zz,BESIII:2018cnd,BESIII:2022qax,ParticleDataGroup:2024cfk,Lee:1957qs}\\


\newpage


\renewcommand{\thefigure}{S\arabic{figure}}
\renewcommand{\thetable}{S\arabic{table}}
\renewcommand{\theequation}{S\arabic{equation}}
\renewcommand{\thepage}{S\arabic{page}}
\setcounter{figure}{0}
\setcounter{table}{0}
\setcounter{equation}{0}
\setcounter{page}{1} 


\begin{center}
\section*{Supplementary Materials for\\ \scititle}

The BESIII Collaboration\\ 
\end{center}

\subsubsection*{This PDF file includes:}
BESIII Collaboration Author List\\
Materials and Methods\\
Fig.~S1 \\
Tables~S1 to S3\\
References~\cite{BESIII:2009fln,BESIII:2020nme,GEANT4:2002zbu,Deng:2006,Jadach:2000ir,Ping:2008zz,Chen:2000tv,Yang:2014vra,Liao:2009zz,BESIII:2018cnd,BESIII:2022qax,ParticleDataGroup:2024cfk,Lee:1957qs}\\

\newpage
\subsection*{BESIII Collaboration authors and affiliations}
M.~Ablikim$^{1}$, M.~N.~Achasov$^{4,c}$, P.~Adlarson$^{77}$, X.~C.~Ai$^{82}$, R.~Aliberti$^{36}$, A.~Amoroso$^{76A,76C}$, Q.~An$^{73,59,a}$, Y.~Bai$^{58}$, O.~Bakina$^{37}$, Y.~Ban$^{47,h}$
% ...
\\
\vspace{0.2cm}
(BESIII Collaboration)\\
{\it
$^{1}$ Institute of High Energy Physics, Beijing 100049, People's Republic of China\\
% ...
$^{4}$ Budker Institute of Nuclear Physics SB RAS (BINP), Novosibirsk 630090, Russia\\
% ...
$^{36}$ Johannes Gutenberg University of Mainz, Johann-Joachim-Becher-Weg 45, D-55099 Mainz, Germany\\
$^{37}$ Joint Institute for Nuclear Research, 141980 Dubna, Moscow region, Russia\\
% ...
$^{47}$ Peking University, Beijing 100871, People's Republic of China\\
% ...
$^{58}$ Southeast University, Nanjing 211100, People's Republic of China\\
$^{59}$ State Key Laboratory of Particle Detection and Electronics, Beijing 100049, Hefei 230026, People's Republic of China\\
% ...
$^{73}$ University of Science and Technology of China, Hefei 230026, People's Republic of China\\
% ...
$^{76}$ University of Turin and INFN, (A)University of Turin, I-10125, Turin, Italy; (B)University of Eastern Piedmont, I-15121, Alessandria, Italy; (C)INFN, I-10125, Turin, Italy\\
$^{77}$ Uppsala University, Box 516, SE-75120 Uppsala, Sweden\\
% ...
$^{82}$ Zhengzhou University, Zhengzhou 450001, People's Republic of China\\
$^{a}$ Deceased\\
% ...
$^{c}$ Also at the Novosibirsk State University, Novosibirsk, 630090, Russia\\
% ...
$^{h}$ Also at State Key Laboratory of Nuclear Physics and Technology, Peking University, Beijing 100871, People's Republic of China\\
% ...

}

\newpage

\subsection*{Materials and Methods}
\subsubsection*{BESIII experiment and simulation}
The BESIII detector~\cite{BESIII:2009fln,BESIII:2020nme} is designed to collect data from electron-positron collisions in the center-of-mass energy range from 1.84 to 4.95 GeV. The cylindrical core of the BESIII detector covers $93\%$ of the full solid angle. The detector includes a helium-based multilayer drift chamber (MDC), providing measurements of the momentum and decay vertex position of particle. Charged particle identification is performed using information from the MDC and a plastic scintillator time-of-flight system.

Simulated data are used to optimize the event selection algorithm, estimate potential background types and levels, improve the signal-to-noise ratio of the final dataset, and accurately compute the normalization factors for the maximum likelihood fit.  The simulation is performed using the GEANT4-based~\cite{GEANT4:2002zbu} BESIII Object Oriented Simulation Tool project~\cite{Deng:2006}, where Geant4 is responsible for describing the geometry of the BESIII detector, the transportation of particles within the detector, and their interactions with matter. The production of $J/\psi$ and known decay processes is simulated using the KKMC~\cite{Jadach:2000ir} model and the BesEvtGen~\cite{Ping:2008zz} model, while Lundcharm~\cite{Chen:2000tv,Yang:2014vra} is responsible for simulating the unknown decay modes of $J/\psi$.
To reduce discrepancies between the experimental and simulated data, a control channel $ J/\psi \rightarrow p\bar{p}\pi^{+}\pi^{-} $ is used to estimate differences in tracking efficiency for charged particles. The discrepancy serves as a correction factor to reweight events in the simulated data, thereby decreasing the systematic uncertainties related to the tracking efficiency. Additionally, the detection efficiency in the simulated data is corrected to account for discrepancies in the azimuthal angle distributions of protons and antiprotons.

\subsubsection*{Selection Criteria}
The data sample used in this analysis is based on $(10.087\pm0.044)\times10^{9} J/\psi$ events collected by BESIII in 2009, 2012, and 2017-2019. The event selection process for the data sample is as follows: $\Lambda$ and $\bar{\Lambda}$ are required to decay to the charged final states $p\pi^{-}$ and $\bar{p}\pi^{+}$, respectively. The main drift chamber of the BESIII detector detects the trajectories of charged particles. Charged tracks detected in the MDC are required to be within a polar angle ($\theta$) range of $|\rm{cos\theta}|<0.93$, where $\theta$ is defined with respect to the $z$-axis,
which is the symmetry axis of the MDC. Charged tracks with momentum greater than $0.5\ \rm{GeV}/c$ are considered proton candidates, while those with momentum below this threshold are considered candidates for pions. Subsequently, the $\Lambda$ and $\bar{\Lambda}$ candidates are constructed by pairing oppositely charged protons and pions with a vertex reconstruction, which constrains the production position of $p\pi^{-}$ and $\bar{p}\pi^{+}$ to a common vertex. 
Due to the long intrinsic lifetime of $\Lambda(~10^{-10}\ s)$, the decay vertex of $\Lambda$ is generally located far from the $e^{+}e^{-}$ collision point. Therefore, by constraining the fitting of the vertex and requiring a distance greater than zero from the $e^{+}e^{-}$ interaction point, background processes originating from the collision point, such as $J/\psi \rightarrow p\bar{p}\pi^{+}\pi^{-}, \Delta^{++}\bar{p}\pi^{-}, \bar{\Delta}^{--}p\pi^{+}$ and $\Delta^{++}\bar{\Delta}^{--}$, can be effectively suppressed.
To further reduce the background contribution from non-$\Lambda$ events such as $J/\psi \rightarrow p\bar{p}\pi^{+}\pi^{-}$, the mass of $\Lambda$ is restricted to be between 1.111 GeV/$c^{2}$ and 1.121 GeV/$c^{2}$.
Finally, an energy–momentum conservation constraint is performed on the \( p\bar{p}\pi^{+}\pi^{-} \) hypothesis, requiring the sum of the four–momentum of the final–state particles to equal the initial four-momentum of \( e^+e^- \). This fit imposes four independent constraints, corresponding to four degrees of freedom, and events are required to satisfy \(\chi^{2} < 60\).
This effectively removes background sources such as \( J/\psi \rightarrow \gamma \eta_{c}(\eta_{c} \rightarrow \Lambda \bar{\Lambda}) \) and \( J/\psi \rightarrow \gamma \Lambda \bar{\Lambda} \), where these background decay modes contain the same final-state decay products as the signal decay mode, but with an additional photon.

After applying all the event selection criteria described above, the final number of events in the data sample is 3049187, with an estimated background of $3801 \pm 63$ events, resulting in a purity of $99.9\%$. 
The number of non-peaking background events in the data is estimated using the sideband region of the two-dimensional invariant mass spectrum of $p \pi^{-}$ and $\bar{p} \pi^{+}$, with the lower and higher sideband regions defined as $m_{p\pi^{-}/\bar{p}\pi^{+}} \in [1.098, 1.107]\ \rm{GeV}/c^{2}$ and $m_{p\pi^{-}/\bar{p}\pi^{+}} \in [1.125, 1.134]\ \rm{GeV}/c^{2}$, respectively. 
The peaking background of \( J/\psi \rightarrow \gamma \eta_{c}\big(\eta_{c} \rightarrow \Lambda \bar{\Lambda}\big) \) and \( J/\psi \rightarrow \gamma \Lambda \bar{\Lambda} \) is estimated using dedicated exclusive Monte Carlo samples and normalized to the expected yields using PDG branching fractions. In the simulation, the decay \( J/\psi \rightarrow \gamma \eta_{c} \) is generated with an angular distribution \( 1 + \cos^{2}\theta_{\gamma} \)~\cite{Liao:2009zz}, where \(\theta_{\gamma}\) is the angle between the photon and the \(e^{+}\) beam direction in the center-of-mass frame; the process \( J/\psi \rightarrow \gamma \Lambda \bar{\Lambda} \) is modeled with a pure phase-space distribution.

\subsubsection*{ Differential decay probability }

\begin{figure}
    \centering
    \includegraphics[width=0.6\textwidth]{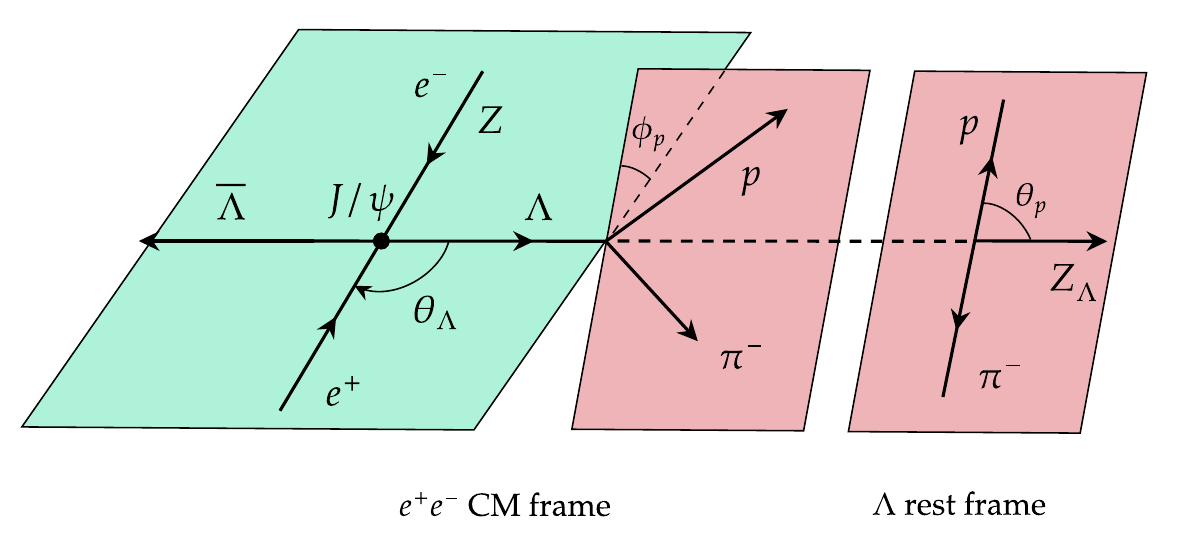}
    \caption{\textbf{Definition of the Cartesian coordinate system for the \(e^{+}e^{-} \rightarrow J/\psi \rightarrow \Lambda  \overline{\Lambda} \) with \(\Lambda ( \overline{\Lambda}) \rightarrow p \pi^{-} (\bar{p} \pi^{+})\).} In the rest frame of \(\Lambda\), the coordinate system \((X_{\Lambda}, Y_{\Lambda}, Z_{\Lambda})\) is defined with \(Z_{\Lambda}\) aligned along the momentum direction of \(\Lambda\) in the \(e^{+}e^{-}\) center-of-mass frame. The other unit vectors are defined as \(Y_{\Lambda} = Z \times Z_{\Lambda}\) and \(X_{\Lambda} = Y_{\Lambda} \times Z_{\Lambda}\), in which $Z$ denotes the momentum direction of the electron beam in the \(e^{+}e^{-}\) center-of-mass frame. For the \( \overline{\Lambda}\), the coordinate system \((X_{ \overline{\Lambda}}, Y_{ \overline{\Lambda}}, Z_{ \overline{\Lambda}})\) is constructed as \((-X_{\Lambda}, Y_{\Lambda}, -Z_{\Lambda})\). }
    \label{fig:coordinate}
\end{figure}

The  differential decay probability of the $e^+e^-\to J/\psi\to\Lambda\overline{\Lambda}\to p\pi^-\overline{p}\pi^+$ process is given by~\cite{Fu:2023ose} 
\begin{equation}\label{angdis}
\begin{aligned}
 \frac{d\sigma}{d\Omega}\propto \sum_{\left[\lambda\right],m,m^{\prime}}&\rho_{m,m^{\prime}}d^{j=1}_{m,\lambda_{\Lambda}-\lambda_{ \overline{\Lambda}}}(\theta_{\Lambda})d^{j=1}_{m^{\prime},\lambda^{\prime}_{\Lambda}-\lambda^{\prime}_{ \overline{\Lambda}}}(\theta_{\Lambda})
\mathcal{M}_{\lambda_{\Lambda},\lambda_{ \overline{\Lambda}}}\mathcal{M}^{*}_{\lambda^{\prime}_{\Lambda},\lambda^{\prime}_{ \overline{\Lambda}}}\delta_{m,m^{\prime}}\\
 &\times D^{*j=1/2}_{\lambda_{\Lambda},\lambda_{p}}(\phi_{p},\theta_{p})D^{j=1/2}_{\lambda^{\prime}_{\Lambda},\lambda^{\prime}_{p}}(\phi_{p},\theta_{p})\mathcal{H}_{\lambda_{p}}\mathcal{H}^{*}_{\lambda^{\prime}_{p}}
 \\ &\times D^{*j=1/2}_{\lambda_{ \overline{\Lambda}},\lambda_{\bar{p}}}(\phi_{\bar{p}},\theta_{\bar{p}})D^{j=1/2}_{\lambda^{\prime}_{ \overline{\Lambda}},\lambda^{\prime}_{\bar{p}}}(\phi_{\bar{p}},\theta_{\bar{p}})\overline{\mathcal{H}}_{\lambda_{\bar{p}}}\overline{\mathcal{H}}^{*}_{\lambda^{\prime}_{\bar{p}}}, \\
\end{aligned}
\end{equation}
where $\left[\lambda\right]$ is a set of all possible indexes representing the helicities of $\Lambda$ and $\overline{\Lambda}$ via the $J/\psi$ resonance and the helicities of $p$ and $\bar{p}$ via $\Lambda$ and $\overline{\Lambda}$, respectively.
Here, $\rho_{m,m'}$ denotes the spin-density matrix of $J/\psi$, while $d$ and $D$ are the Wigner small \(d\)-matrix and the full \(D\)-matrix. 
The longitudinal polarization of the $J/\psi$ meson, $P_{L}$, is characterized by the difference between the diagonal density matrix elements, $\rho_{++}$ and $\rho_{--}$, normalized to their sum.  
In experiments such as BESIII, where the incoming beams are unpolarized, $P_{L}$ is directly related to the left–right asymmetry $\mathcal{A}_{LR}^{0}$:
\begin{equation}
\begin{aligned}
P_{L} \equiv \mathcal{A}_{LR}^{0} = 
\frac{\sigma_{R}-\sigma_{L}}{\sigma_{R}+\sigma_{L}} 
= \frac{-\sin^{2}\theta_{W}^{\text{eff}}+3/8}
{2\sin^{2}\theta_{W}^{\text{eff}}\cos^{2}\theta_{W}^{\text{eff}}}
\frac{m_{J/\psi}^{2}}{m_{Z}^{2}}.
\end{aligned}
\end{equation}
Here, \(\sigma_{R}\) and \(\sigma_{L}\) denote the production cross sections of \(J/\psi\) with right- and left-handed electrons, respectively, and \(m_Z\) represents the mass of the \(Z\) boson.  
This asymmetry arises from the effective weak mixing angle \(\theta_W^{\mathrm{eff}}\), with a predicted magnitude on the order of \(10^{-4}\)~\cite{He:2022jjc}.
The quantities $\mathcal{M}_{\lambda_{\Lambda},\lambda_{\overline{\Lambda}}}$ represent the helicity amplitudes for the production process $J/\psi\to\Lambda\overline{\Lambda}$, encoding the relative strengths and phases of the allowed helicity configurations. The decay dynamics of the hyperons is described by the helicity amplitudes $\mathcal{H}_{\lambda_p}$ for $\Lambda\to p\pi^-$ and $\overline{\mathcal{H}}_{\lambda_{\bar{p}}}$ for $\overline{\Lambda}\to\overline{p}\pi^+$. 
The direction of $\Lambda$ in the $J/\psi$ rest frame, along with the motion directions of the proton ($p$) and antiproton ($\bar{p}$) in the rest frames of the $\Lambda$ and $\overline{\Lambda}$, respectively, are described by the kinematic variable  $\xi = (\theta_{\Lambda}, \theta_{p}, \theta_{\bar{p}}, \phi_{p}, \phi_{\bar{p}})$.
The definition of the corresponding coordinate system, shown in Fig.~\ref{fig:coordinate}, is as follows:  
In the rest frame of the $\Lambda$ hyperon, the Cartesian coordinate system ($X_{\Lambda}, Y_{\Lambda}, Z_{\Lambda}$) is defined, with the unit vector $Z_{\Lambda}$ aligned with the momentum direction of $\Lambda$ in the center-of-mass frame for $e^{+}e^{-}$. Unit vectors $Y_{\Lambda}$ and $X_{\Lambda}$ are defined as $Y_{\Lambda}=Z \times Z_{\Lambda}$ and $X_{\Lambda}=Y_{\Lambda} \times Z_{\Lambda}$, respectively. For the $\bar{\Lambda}$, the coordinate system ($X_{\bar{\Lambda}}, Y_{\bar{\Lambda}}, Z_{\bar{\Lambda}}$) is constructed as ($-X_{\Lambda}, Y_{\Lambda}, -Z_{\Lambda}$). 
The motion directions of the proton $p$ and antiproton $\bar{p}$ are characterized by the angles ($\theta_{p}, \theta_{\bar{p}}, \phi_{p}, \phi_{\bar{p}})$ in their separate systems.

The amplitude for the $J/\psi\to\Lambda\overline{\Lambda}$ decay is written as~\cite{He:2022jjc} 
\begin{equation}
	\begin{aligned}
		\mathcal{M}_{\lambda_{\Lambda},\lambda_{\overline{\Lambda}}}=\epsilon_{\mu}(\lambda_{\Lambda}-\lambda_{\overline{\Lambda}})\bar{u}(\lambda_{\Lambda},p_{\Lambda}) (F_{V}\gamma^{\mu}+\frac{i}{2m_{\Lambda}}\sigma^{\mu\nu}q_{\nu}H_{\sigma}
		+\gamma^{\mu}\gamma^{5}F_{A}+\sigma^{\mu\nu}\gamma^{5}q_{\nu}H_{T} )v(\lambda_{\overline{\Lambda}},p_{\overline{\Lambda}}),
	\end{aligned}
\end{equation}
where $m_{\Lambda}$ is $\Lambda$ hyperon mass, and $p_{\Lambda}$ and $p_{\overline{\Lambda}}$ are the four momenta of $\Lambda$ and $\overline{\Lambda}$, respectively.
The form factors $F_V$ and $H_{\sigma}$ are related to the electric and magnetic form factors $G_{2}$ and $G_{1}$~\cite{He:2022jjc}, with $F_{V}=G_{1}-4m^{2}_{\Lambda}\frac{(G_{1}-G_{2})}{(p_{\Lambda}-p_{\overline{\Lambda}})^{2}}$ and $H_{\sigma}=4m^{2}_{\Lambda}\frac{(G_{1}-G_{2})}{(p_{\Lambda}-p_{\overline{\Lambda}})^{2}}$.
Two parameters are introduced to characterize the $\Lambda$ polarization:  
\begin{equation}
    \begin{aligned}
\alpha_{J/\psi} = \frac{m^{2}_{J/\psi}\left|G_{1}\right|^{2} - 4m^{2}_{\Lambda}\left|G_{2}\right|^{2}}
                       {m^{2}_{J/\psi}\left|G_{1}\right|^{2} + 4m^{2}_{\Lambda}\left|G_{2}\right|^{2}}, 
\quad 
\frac{G_{1}}{G_{2}} = \left|\frac{G_{1}}{G_{2}}\right| e^{-i\Delta\Phi},
    \end{aligned}
\end{equation}
where $\Delta\Phi$ denotes the relative phase between the form factors $G_{1}$ and $G_{2}$~\cite{BESIII:2018cnd,BESIII:2022qax}.
The form factor $G_{1} = (161 \pm 4)\times10^{-5}$, determined from the measured decay rate of $J/\psi\to\Lambda\overline{\Lambda}$ and the parameter $\alpha_{J/\psi}$~\cite{Du:2024jfc, ParticleDataGroup:2024cfk}, is treated as a real parameter and fixed as a constant in the subsequent fit.
The helicity amplitudes $\mathcal{H}_{\lambda_{p}}$ and $\overline{\mathcal{H}}_{\lambda_{\bar p}}$ describe the weak decays $\Lambda\to p\pi^-$ and $\overline{\Lambda}\to \overline{p}\pi^+$, from which the decay-asymmetry parameters $\alpha_{\Lambda}$ and $\alpha_{\overline{\Lambda}}$ are defined. Their combination yields the CP-violating observable $
A_{CP} = \frac{\alpha_{\Lambda}+\alpha_{\overline{\Lambda}}}{\alpha_{\Lambda}-\alpha_{\overline{\Lambda}}}$, 
which quantifies the relative difference between $\Lambda$ and $\overline{\Lambda}$ decays~\cite{Lee:1957qs}.
A non-zero transverse polarization of $\Lambda$, along with the most precise test of CP symmetry conservation in $\Lambda$ decay, has recently been measured at BESIII~\cite{BESIII:2022qax}.

The parameter \(g_{V}\) is a non-perturbative quantity, analogous to a decay constant, and can be defined through the following matrix element:
\begin{equation}
\langle 0| \bar{c} \gamma^{\mu} c |J/\psi \rangle = g_{V} \, \epsilon^{\mu} ,
\end{equation}
where \(\epsilon^{\mu}\) is the polarization vector of the \(J/\psi\) meson. In this study, we focus on the leptonic decay channel \(J/\psi \to e^{+} e^{-}\), which allows for a clean extraction of \(g_{V}\) from the decay amplitude. The corresponding matrix element for this process can be written as
\begin{equation}
\mathcal{M}_{\lambda_{1},\lambda_{2}} = \frac{2}{3} g_{V} \, \epsilon^{\mu}(\lambda_{1}-\lambda_{2}) \frac{-i g_{\mu\nu}}{m_{J/\psi}^{2}} \, \bar{u}(k,\lambda_{1}) (-i e \gamma^{\nu}) v(-k,\lambda_{2}) ,
\end{equation}
where \(k\) denotes the momentum of the outgoing electron, and \(\lambda_{1,2}\) are the helicities of the electron and positron. The factor \(2/3\) arises from the charm quark electric charge in units of \(e\).

Evaluating the dominant helicity configuration, one obtains
\begin{equation}
\mathcal{M}_{-1/2,+1/2} = \frac{2}{3} e g_{V} \frac{1}{m_{J/\psi}^{2}} 2 \sqrt{2} E = \frac{2 \sqrt{2}}{3 m_{J/\psi}} e g_{V} ,
\end{equation}
where \(E\) is the energy of the electron in the rest frame of the \(J/\psi\). The other significant helicity amplitude, \(\mathcal{M}_{+1/2,-1/2}\), is related by parity, while amplitudes such as \(\mathcal{M}_{+1/2,+1/2}\) are suppressed by the small ratio \(m_{e}/m_{J/\psi}\) and can be safely neglected.

The partial decay width for the leptonic channel is then determined from the squared amplitude through
\begin{equation}
\Gamma(J/\psi \to e^{+} e^{-}) = \frac{(2\pi)^{4}}{2 m_{J/\psi}} \frac{1}{4 (2\pi)^{5}} \left| \mathcal{M} \right|^{2} ,
\end{equation}
with
\begin{equation}
\left| \mathcal{M} \right|^{2} = \frac{1}{3} \left[ 2 \times \frac{4}{9 m_{J/\psi}^{2}} 2 e^{2} g_{V}^{2} \right] ,
\end{equation}
where the factor \(1/3\) accounts for averaging over the three polarization states of the \(J/\psi\) meson.

From these relations, the vector decay constant can be expressed as
\begin{equation}\label{gv}
g_{V}^{2} = \frac{27 \pi \, \Gamma \, m_{J/\psi}^{3}}{e^{2}} = \frac{27 \pi \, \Gamma_{J/\psi} \, \mathcal{BR}(J/\psi \to e^{+} e^{-}) \, m_{J/\psi}^{3}}{e^{2}} ,
\end{equation}
where \(\Gamma_{J/\psi}\) is the total decay width and \(\mathcal{BR}(J/\psi \to e^{+} e^{-})\) is the branching ratio for the leptonic decay.

For the numerical evaluation, we adopt the experimental values from the Particle Data Group~\cite{ParticleDataGroup:2024cfk}:
\begin{equation}
\Gamma_{J/\psi} = 92.6 \pm 1.7~\text{keV} = (92.6 \pm 1.7) \times 10^{-6}~\text{GeV}, \quad
\mathcal{BR}(J/\psi \to e^{+} e^{-}) = (5.971 \pm 0.032)\% .
\end{equation}
Substituting these inputs into Eq.~\ref{gv} yields
\begin{equation}
g_{V} = (1.382 \pm 0.013)~\text{GeV}^{2} .
\end{equation}

\subsubsection*{Maximum log-likelihood fit procedure}
A maximum likelihood fit is performed on the data based on the angular distribution defined as:
\begin{eqnarray}                                                               
        \mathcal{L} = \prod^{N}_{i=1}\mathcal{P}(\xi^i;\vec{\lambda}) =            
        \prod^{N}_{i=1}\mathcal{C}\mathcal{W}(\xi^i;\vec{\lambda})\epsilon(\xi^i).
    \label{eq:ML_likelihood}                      
\end{eqnarray}                                                                 
Here, $\mathcal{P}(\xi^i;\vec{\lambda})$ is the probability density function where the set of helicity angles $\xi^i = (\theta_{\Lambda}^i, \theta_{p}^i, \theta_{\bar{p}}^i, \phi_{p}^i, \phi_{\bar{p}}^i)$ denotes the event $i$, and $\vec{\lambda} = (P_{L}, G_{2}, F_{A}, H_{T}, \alpha_{\Lambda}$, $\alpha_{ \overline{\Lambda}})$ represents the fitted parameters. The joint angular distribution $\mathcal{W}(\xi^i;\vec{\lambda}) \propto \frac{d\sigma}{d\Omega}$ is defined in Eq.~\ref{angdis}.
The detection efficiency is denoted by $\epsilon(\xi^i)$. The normalization factor $\mathcal{C}^{-1} = \frac{1}{N_{\rm{MC}}}\sum^{N^{\prime}}_{j=1}\mathcal{W}(\xi^j;\vec{\lambda})\epsilon(\xi^j)$ is estimated using a simulated data of \( N' \) events generated uniformly in the phase space, selected based on the same event criteria as in the data analysis. Simulated data are generated at approximately 100 times the size of the accepted data to improve the accuracy of the normalization factor. 
The fit is performed using \texttt{RooFit} package~\cite{Verkerke:2003ir}, implementing an unbinned maximum-likelihood fit of the full angular-distribution probability density function. The objective function is defined as
\begin{eqnarray}
\mathcal{S} = -\ln \mathcal{L}_{\rm data} + \ln \mathcal{L}_{\rm bg},
\end{eqnarray}
where $\mathcal{L}_{\rm data}$ and $\mathcal{L}_{\rm bg}$ are the likelihood values for the data and the background events, respectively. The normalization of the PDF is computed via phase-space Monte Carlo integration. Minimization is carried out using \texttt{MINUIT/MIGRAD}, and the parameter uncertainties as well as the full covariance matrix are obtained using \texttt{HESSE}. For weighted Monte Carlo samples, the \texttt{SumW2Error} option is applied to correctly propagate statistical weights. The fit results are reported in Table~\ref{tab:fit_results_all}, and are found to be consistent with our previously published measurements~\cite{BESIII:2022qax} within uncertainties.
Parameters derived from combinations of fitted parameters take into account their correlations, and the full set of correlation coefficients is given in Table~\ref{tab:coeff}.

\subsubsection*{Systematic uncertainties}

The systematic uncertainties in this measurement arise from various sources, including background estimation, charged particle reconstruction, vertex reconstruction for \(\Lambda (\overline{\Lambda})\), energy-momentum conservation constraints, detector resolution, and the values of the parameters \( G_1 \), \( m_{J/\psi} \), \( m_Z \), and \( g_V \). Systematic effects related to charged particle reconstruction, vertex reconstruction, and energy-momentum conservation constraints are investigated using data-driven methods with different control samples to quantify the discrepancies between experimental and simulated data. Table~\ref{tab:sys_summary} provides a summary of the absolute systematic uncertainties from each source. The total systematic uncertainty for each fitting parameter is obtained by combining the individual contributions in quadrature.

\begin{enumerate}

\item \textbf{Background estimation.}
The systematic uncertainties associated with the background originate from uncertainties in estimating the number of background events. To evaluate this effect, the number of background events is varied by one standard deviation in the joint maximum-likelihood fit. The resulting deviation in the fitted parameters is taken as the systematic uncertainty. Studies show that the impact of this uncertainty on the fitting parameters is negligible.

\item \textbf{Charged particle reconstruction.} 
The systematic uncertainty from the reconstruction efficiency of charged particles is studied using the control sample \( J/\psi \rightarrow p \bar{p} \pi^{+} \pi^{-} \). This sample is used to evaluate the discrepancy in track reconstruction efficiency for charged particles (\( p, \bar{p}, \pi^{+}, \pi^{-} \)) in the MDC between experimental and simulated data. It is assumed that the tracking efficiency depends only on the polar angle \( \theta \) and transverse momentum \( P_T \) of the charged particles. To account for this, the control sample is divided into two-dimensional bins in \( (P_T, \cos\theta) \), and the reconstruction efficiencies in experimental and simulated data are calculated separately in each bin. The ratio of these efficiencies is then used to define a correction factor, which is applied event by event to the simulated data so that the tracking efficiencies match those observed in the experimental data.
The systematic uncertainty in charged particle reconstruction arises from statistical fluctuations in the efficiency correction factor applied to the simulated data. To quantify this uncertainty, the correction factor’s central value is sampled from a Gaussian distribution, incorporating its statistical uncertainty. This procedure is repeated 100 times, generating 100 sets of simulated data with varying correction factors for the maximum likelihood fit. The resulting 100 sets of fit results form a distribution, whose standard deviation is taken as the systematic uncertainty. This uncertainty was found to be negligible.

\item \textbf{$\Lambda$ and $\bar{\Lambda}$ vertex reconstruction.} 
The systematic uncertainty from the efficiency of the \(\Lambda\) and \(\bar{\Lambda}\) vertex reconstruction is evaluated using the control sample \( J/\psi \rightarrow \Lambda \bar{\Lambda}, \Lambda(\bar{\Lambda}) \rightarrow p\pi^{-}(\bar{p}\pi^{+}) \). To account for possible kinematic dependencies, the control sample is divided into \(5\times 5\times 5\times 5\) bins according to the polar angles of the final-state particles \( p, \bar{p}, \pi^{+}, \pi^{-} \), and the efficiency difference between experimental and simulated data is calculated in each bin. The nominal fit results are obtained using the uncorrected simulated data, while the corrected sample is only used to evaluate the potential impact of the efficiency difference. The systematic uncertainty due to the vertex fit is then defined as the difference between the fitting results with and without this correction. Following the methodology used for charged particle reconstruction uncertainties, the correction is applied 100 times to generate 100 sets of simulated data with varying correction factors for the fitting. The systematic uncertainty for each parameter is determined as the difference between the central values of these 100 fits and the measured results.

\item \textbf{Energy-momentum conservation constraints.}
The systematic uncertainty from the efficiency of the energy-momentum conservation constraints is evaluated using the control sample \( J/\psi \rightarrow \Lambda \bar{\Lambda}, \Lambda(\bar{\Lambda}) \rightarrow p\pi^{-}(\bar{p}\pi^{+}) \). 
To probe potential discrepancies between experimental and simulated datas, the control sample is partitioned into a four-dimensional distribution with \(10 \times 10 \times 10 \times 10 \) bins over the polar angles \(\theta_{p}, \theta_{\bar{p}}, \theta_{\pi^{+}}, \theta_{\pi^{-}}\). 
Within each bin, the simulated efficiency is corrected to match that of the data. 
The nominal results are obtained from the uncorrected simulated data, while the systematic effect is evaluated by applying the bin-by-bin correction and assigning the resulting shift in the fit parameters as the uncertainty from the energy–momentum conservation constraints. To test the stability of this estimate, the bin-by-bin correction factors are randomly varied within their statistical uncertainties and propagated through 100 repeated fits; the systematic uncertainty for each parameter is then taken as the difference between the measured value and the central value obtained from these 100 fits.

\item \textbf{Detector resolution.}
To assess the impact of detector resolution, we generated 10 sets of simulated data, each with the same size as the experimental data. These datasets were sampled according to the angular distribution in Equation~\ref{angdis}, with input parameters fixed to the measured value obtained from the experimental data. The samples were reconstructed using the same event selection criteria as applied to the experimental data. After fitting the ten sets of simulated data, the differences between the central values of the fitting results and the input values were taken as the systematic uncertainty.

\item \textbf{The value of $G_{1}$.}
Since all form factors were normalized to \( G_{1} \) during the maximum likelihood fit, the final fitting results of these form factors are influenced by the magnitude of \( G_{1} \). Given that \( G_{1} = (161 \pm 4) \times 10^{-5} \), the relative systematic uncertainty introduced by these form factors is estimated to be \( 2.5\% \).

\item \textbf{The value of $m_{J/\psi}$, $m_{Z}$ and $g_V$.}
The input values of \( m_{J/\psi} \) and \( m_{Z} \) are required to calculate \( \sin^{2}\theta_{W} \). According to the latest results from \cite{ParticleDataGroup:2024cfk}, the relative systematic uncertainty introduced by \( m_{J/\psi} \) and \( m_{Z} \) on \( \sin^{2}\theta_{W} \) is estimated to be \( 0.2\% \), which is negligible. Additionally, the relative systematic uncertainty introduced by \( g_V \) on \( \text{Re}(d_{\Lambda}) \) and \( \text{Im}(d_{\Lambda}) \) is estimated to be \( 0.9\% \), which is also negligible.

\end{enumerate}
\newpage

\newpage
\clearpage

\begin{table}[htbp]
    \centering
    \caption{
    \textbf{Summary of fitted and derived parameters.} 
The fit yields the $J/\psi$ polarization $P_{L}$ and the decay form factors $G_{2}$, $F_{A}$, and $H_{T}$ in $J/\psi$ decays, together with the decay asymmetry parameters $\alpha_{\Lambda}$ ($\alpha_{\overline{\Lambda}}$) in $\Lambda \to p\pi^{-}$ ($\overline{\Lambda} \to \overline{p}\pi^{+}$). From these results we derive the $\Lambda$ polarization observables $\alpha_{J/\psi}$ and $\Delta\phi$, the effective weak mixing angle $\sin^{2}\theta_{W}^{\text{eff}}$, the CP-violating asymmetry $A_{CP}$ in $\Lambda$ decays, and the $\Lambda$ electric dipole moment $d_{\Lambda}$. Uncertainties are quoted with the statistical component first, followed by the systematic.}
    \scalebox{1.}{
        \begin{tabular}{lll}
            \\
            \hline
            \hline
            \textbf{\textrm{Parameter}} & \textbf{\textrm{This measurement}} & \textbf{\textrm{Previous measurement}}~\cite{BESIII:2022qax}  \\ 
            \hline 
            $\alpha_{\Lambda}$             & $0.7524 \pm 0.0036 \pm 0.0008 $ & $0.7519 \pm 0.0036 \pm 0.0024 $  \\ \hline 
            $\alpha_{\overline{\Lambda}}$  & $-0.7571 \pm 0.0036 \pm 0.0008 $ & $-0.7559 \pm 0.0036 \pm 0.0030$ \\ \hline 
            $Re(G_{2})$  & $(9.71  \pm 0.06 \pm 0.24) \times 10^{-4}$ & -   \\ \hline 
            $Im(G_{2})$  & $(9.14  \pm 0.04 \pm 0.23) \times 10^{-4}$ & -   \\ \hline 
            $P_{L}$      & $(-1.8  \pm 1.2 \pm 0.8)   \times 10^{-3}$ & -   \\ \hline 
            $Re(F_{A})$  & $(-2.4  \pm 1.6 \pm 3.1)   \times 10^{-6}$ & -   \\ \hline 
            $Im(F_{A})$  & $(-7.9  \pm 3.7 \pm 2.5)   \times 10^{-6}$ & -  \\ \hline 
            $Re(H_{T})$  & $(-1.4  \pm 1.4 \pm 0.2)   \times 10^{-6}$ $\textrm{GeV}^{-1}$ & -   \\ \hline 
            $Im(H_{T})$  & $(1.3   \pm 1.2 \pm 0.4)   \times 10^{-6}$ $\textrm{GeV}^{-1}$ & -   \\  
            \hline
            \hline
            $\alpha_{J/\psi}$  & $0.4748 \pm 0.0022 \pm 0.0017$ & $0.4748 \pm 0.0022 \pm 0.0031$  \\ \hline 
            $\Delta\phi$  & $0.7552 \pm 0.0042 \pm 0.0013$ & $0.7521 \pm 0.0042 \pm 0.0066$ \\ \hline 
            $A_{CP}$  & $(-3.1 \pm 4.6 \pm 1.1) \times 10^{-3}$ & $(-2.5 \pm 4.6 \pm 1.2) \times 10^{-3}$ \\ \hline 
            $\sin^{2}\theta_{W}^{\text{eff}}$  & $-0.15 \pm 0.12 \pm 0.26$ & -   \\ \hline 
            $Re(d_{\Lambda})$  & $(-3.1 \pm 3.2 \pm 0.5) \times 10^{-19}$ $e\ \textrm{cm}$ & -   \\ \hline
            $Im(d_{\Lambda})$  & $(2.9 \pm 2.6 \pm 0.6) \times 10^{-19}$ $e\ \textrm{cm}$ & -   \\ \hline
            \hline
        \end{tabular}
    }
    \label{tab:fit_results_all}
\end{table}

\newpage
\clearpage

\begin{table}[htbp]
  \centering
  \caption{\textbf{Absolute systematic uncertainties on the fitted parameters.}}
  
  \scalebox{0.9}{
    \begin{tabular}{ccccc|c}
        \\
        \hline
        \hline
        \textbf{Source} & \textbf{$\Lambda/ \overline{\Lambda}$ vertex fit} 
        & \textbf{Kinematic fit} & \textbf{Detector resolution} & \textbf{$G_{1}$ value} & \textbf{Total} \\ 
    \hline 
    $\alpha_{\Lambda} (10^{-3})$        & 0.1 & 0.4  & 0.7 & -   & 0.8 \\ \hline 
    $\alpha_{ \overline{\Lambda}} (10^{-3})$  & 0.1 & 0.2  & 0.8 & -   & 0.8 \\ \hline 
    $Re(G_{2}) (10^{-6})$               & 0.9 & 0.3  & 1.0 & 24  & 24  \\ \hline 
    $Im(G_{2}) (10^{-6})$               & 0.2 & 0.2  & 3.0 & 23  & 23  \\ \hline 
    $P_{L} (10^{-3})$                   & 0.2 & 0.1  & 0.8 & -   & 0.8 \\ \hline 
    $Re(F_{A}) (10^{-6})$               & 0.3 & 0.2  & 3.1 & 0.1 & 3.1 \\ \hline 
    $Im(F_{A}) (10^{-6})$               & 0.1 & 0.4  & 2.5 & 0.2 & 2.5 \\ \hline 
    $Re(H_{T}) (10^{-6}\ \textrm{GeV}^{-1})$               & 0.1 & 0.1  & 0.1 & 0.0 & 0.2 \\ \hline 
    $Im(H_{T}) (10^{-6}\ \textrm{GeV}^{-1})$               & 0.2 & 0.1  & 0.3 & 0.0 & 0.4 \\ 
    \hline 
    \hline 
    $\alpha_{J/\psi} (10^{-3})$         & 0.5 & 0.0  & 1.6 & -   & 1.7 \\ \hline 
    $\Delta\phi (10^{-3})$              & 0.4 & 0.3  & 1.2 & -   & 1.3 \\ \hline 
    $A_{CP} (10^{-3})$                  & 0.0 & 0.1  & 1.0 & -   & 1.1 \\ \hline 
    $sin^{2}\theta_{W} (10^{-1})$       & 0.1 & 0.0  & 2.6 & -   & 2.6 \\ \hline 
    $Re(d_{\Lambda}) (10^{-19}\ e\ \textrm{cm})$        & 0.3 & 0.3  & 0.2 & 0.1 & 0.5 \\ \hline 
    $Im(d_{\Lambda}) (10^{-19}\ e\ \textrm{cm})$        & 0.3 & 0.2  & 0.5 & 0.1 & 0.6 \\ \hline 

    \hline
    \hline
    \end{tabular}
}
  \label{tab:sys_summary}
\end{table}

\newpage
\clearpage

\begin{table}[h] 
       \centering 
       \caption{\textbf{Correlation matrix of the fitted parameters.}} 
       \label{tab:coeff} 
       \begin{tabular}{cccccccccc} 
       \hline 
       \hline
       \textrm{Paras.} & $Im(F_{A})$ & $Im(H_{T})$ & $Im(G_{2})$ & $P_{L}$ & $Re(F_{A})$ & $Re(H_{T})$ & $Re(G_{2})$ 
       & $\alpha_{\Lambda}$ & $\alpha_{\bar{\Lambda}}$ \\ 
       \hline 
       $Im(F_{A})$              &  1.000 &  0.001 &  0.033 & -0.002 & -0.085 & -0.000 &  0.057 &  0.013 & -0.010 \\  
       $Im(H_{T})$              &        &  1.000 &  0.001 & -0.012 & -0.003 &  0.351 & -0.003 & -0.046 & -0.045 \\
       $Im(G_{2})$              &        &        &  1.000 & -0.002 & -0.045 &  0.002 & -0.407 & -0.034 &  0.049 \\
       $P_{L}$                  &        &        &        &  1.000 & -0.000 & -0.029 &  0.007 &  0.023 &  0.019 \\
       $Re(F_{A})$              &        &        &        &        &  1.000 &  0.003 & -0.139 & -0.002 & -0.008 \\
       $Re(H_{T})$              &        &        &        &        &        &  1.000 &  0.003 & -0.039 & -0.038 \\
       $Re(G_{2})$              &        &        &        &        &        &        &  1.000 &  0.097 & -0.100 \\
       $\alpha_{\Lambda}$       &        &        &        &        &        &        &        &  1.000 &  0.848 \\
       $\alpha_{\bar{\Lambda}}$ &        &        &        &        &        &        &        &        &  1.000 \\
       \hline 
       \hline
       \end{tabular} 
   \end{table}


\clearpage 

\end{document}